\documentclass[12pt,a4paper]{article}
\usepackage{authblk}
\usepackage{tikz-cd}
\usepackage[english]{babel}
\usepackage[latin1]{inputenc}
\usepackage{lmodern}
\usepackage{verbatim}
\usepackage{amsfonts}
\usepackage{amsmath}
\usepackage[T1]{fontenc}
\usepackage{color}
\usepackage[normalem]{ulem}
\usepackage{todonotes}
\usepackage{braket}
\usepackage{natbib}
\usepackage{url}
\usepackage[bookmarksnumbered=true, colorlinks=true, citecolor=blue]{hyperref}
\date{}

\newcommand{\rev}[1]{\textcolor{black}{#1}}
\newcommand{\rom}[1]{\textcolor{black}{#1}}

\title{The metaphysics of decoherence}
\author{Antonio Vassallo}
\affil{Faculty of Administration and Social Sciences\\Warsaw University of Technology\\Plac Politechniki 1\\ 00-661 Warsaw, Poland\\antonio.vassallo1977@gmail.com}
\author{Davide Romano}
\affil{Center of Philosophy\\University of Lisbon\\Alameda da Universidade\\1600-214 Lisbon, Portugal\\davideromano1984@libero.it}

\begin{document}

\maketitle

\begin{center}
Forthcoming in \emph{Erkenntnis}.
\end{center}
\pdfbookmark[1]{Abstract}{abstract}
\begin{abstract}
The paper investigates the type of realism that best suits the framework of decoherence taken at face value without postulating a plurality of worlds, or additional hidden variables, or non-unitary dynamical mechanisms. It is argued that this reading of decoherence leads to an extremely radical type of perspectival realism, especially when cosmological decoherence is considered.
\end{abstract}

\section{Reshaping metaphysical realism: The case of quantum mechanics}\label{sec:intro}

\emph{Scientific realism} is the view that science, and physics in particular, tells us a fairly accurate story about reality. Hence, a precise characterization of scientific realism should spell out how scientific theories relate to the natural world. A standard definition of scientific realism, such as that given in \citet[][section 1.2]{664}, characterizes this view as the union of three different components (or theses). The first is dubbed \emph{epistemic} and amounts to the claim that scientific theories provide us with knowledge about how the world is. The second is the \emph{semantic} thesis, which amounts to taking scientific claims at face value, that is, scientific claims are statements about facts obtaining in the world. Finally, there is the \emph{metaphysical} component, which deals with \emph{what it is} for the just mentioned scientific claims to be true. Otherwise said, the metaphysical component deals with what the ontology of the world is according to scientific theories. Here we will focus on this third component (see also \citealp{650} for a discussion of all three theses).

In general, \emph{metaphysical realism} is the view that there are ``things'' which exist independently of any mind that might perceive them. Here we use the word ``thing'' as a placeholder that can be replaced with more precise designations, depending on one's own ontological tastes. Think of material bodies, abstract objects, facts, events, space, time, patterns or other less intuitive entities: all of these things can be part of this mind-independent reality. Although with a varying terminology, there is a general consensus in the literature to broadly characterize metaphysical realism in such a way; for example \citet[][p. 19]{648} calls it ``deep realism'', while for \cite{649} it is ``generic realism''. Clearly, a scientific realist holds the firm belief that science can and should help in clarifying and sharpening the metaphysical realist thesis. The present paper will indeed be concerned with the impact that modern physics has on the notion of metaphysical realism.

In order to clarify and sharpen the metaphysical realist thesis \emph{qua} component of scientific realism, we can start by considering classical mechanics. In this context, the values of the dynamical variables that describe a system have a definite value at any point in time. Furthermore there is a unique physically relevant description of any system, which is the one that can be attained by observers moving inertially. This characterization of classical mechanics---distilled from the contemporary textbook presentations of the topic---seems to force at least two minimal constraints on reality. We will call these requirements \emph{definiteness} and \emph{objectivity}. 

The requirement of definiteness roughly means that all things possess a definite mode of being. For example, if a material object possesses a certain property, such a property has to be definite (or determinate). A colored balloon cannot be \emph{just} colored, it has to have some definite color, say, red. Of course, the balloon might not have been red earlier (it might have been painted red just a moment ago), but even at that point it must have possessed \emph{some} determinate color. 

The requirement of objectivity, instead, states that for any thing, its mode of being is unique. A thing cannot be, say, a red balloon \emph{and} a black horse. A thing might perhaps \emph{appear} differently if perceived under different conditions but, from a purely ontological point of view, the thing \emph{is} in a unique way. This is not to say that a thing cannot be ontologically parasitic on other things---one might argue that a balloon can be reduced to a bunch of polymers holding up a bunch of helium atoms. Rather, this means that the way such a thing is---including how it is possibly made up of other things---has no ontological ambiguity. Two remarks are in order at this point. First, objectivity as we cast it is primarily a metaphysical notion, rather than an epistemic or a semantic one. Certainly, the notion does have epistemic and semantic counterparts, especially in the scientific realist context (e.g. scientific knowledge is objective in that it is knowledge of the world \emph{as is}). However, what we will call objectivity throughout the text is an attribute of reality \emph{simpliciter}: it concerns the way things are. Second, this notion of objectivity plays a pivotal role in standard scientific realism because it helps accounting for the explanatory power of scientific theories (see \citealp{699}, for an overview on the topic of scientific objectivity).

We define the above metaphysical characterization of reality as \emph{standard realism}. Note that standard realism as we defined it says nothing about what particular things inhabit the world and how they behave. For this reason, it is not to be confused with the brand of ``commonsensical'' metaphysical theorizing attacked in \citet[][cf. for example section 1.2]{407}. What Ladyman and Ross attack is the excessive reliance on intuitions and/or outdated physics in shaping the \emph{ontology} of the physical world, which then results in the attempt at shoehorning a fundamental picture of ``microbangings'' between ``little things'' into modern physics. However, these authors are in favor of adopting definiteness and objectivity in the metaphysical inquiry, as their declaration of intent makes clear:
\begin{quote}
\rev{[W]e will articulate a theory of what it is for a scientific theory to be taken to describe a part of \emph{objective reality}---a `real pattern', as we will say [...]\\
(\emph{ibid}, p. 36, emphasis added).}
\end{quote}

An obvious consequence of standard realism as we characterize it is that reality, i.e. the way all existing things are, is metaphysically coherent. This means that reality cannot be decomposed into mutually inconsistent pictures: all these pictures must seamlessly blend to form a consistent whole. Otherwise said, according to standard realism there is an ``outside world'' whose ontological unity is warranted by the existence of definite and objective facts.

At this point a pressing question presents itself: is standard realism, intended as the metaphysical component of scientific realism, still tenable under the light of quantum mechanics? The question is pressing because quantum mechanics is one of the most empirically corroborated theories we have so far, so it is fair to ask how the world is according to this framework.


By ``quantum mechanics'' we mean the formalism presented and discussed in the standard textbooks on the subject, such as \citet{627} and \citet{332}. This theoretical framework relies on a number of basic postulates revolving around the measurement procedure of quantum systems. In a nutshell, the physical state of a system is completely represented by a state vector (or a wave function) in the Hilbert space.\footnote{Throughout the text we will use the terms ``quantum state'', ``state vector'', and ``wave function'' interchangeably, since blurring their distinction does not affect our discussion.} The wave function's dynamics is described by the unitary and linear \emph{Schr\"odinger equation}. In particular, the linearity of the Schr\"odinger equation implies that any linear superposition of its solutions is also a solution. The physical properties of a quantum system are called \emph{observables} and mathematically represented by Hermitian operators defined on the system's Hilbert space. The numerical values of these physical magnitudes are the \emph{eigenvalues} of such operators. When a measurement is performed on a system being in a superposition of eigenstates of the measured observable, the state suddenly ``collapses'' in one of such eigenstates with a probability given by the so-called \emph{Born's rule}.



Quantum states are radically different from classical ones. According to the formalism of quantum mechanics, they can be in a superposition of different and mutually incompatible states (at least, until a measurement is performed on the system). Furthermore, the values of the observables associated to non-commuting operators (e.g., two operators $\hat{A}$ and $\hat{B}$ for which $\hat{A}\hat{B}\neq\hat{B}\hat{A}$) are sensitive to the measurement order. For example, if we perform a measurement of the spin components of an electron first in the $x$-direction and then in the $z$-direction, the final results will be different from what we would have obtained if we performed a measurement in the $z$-direction first and then in the $x$-direction. This is clearly in stark contrast with the classical picture of the electron possessing a definite value of the spin in both directions at the same time. Already at this stage, it seems that quantum mechanics breaks with standard realism and, in particular, with the requirement of ontic definiteness. Indeed, some of the founding fathers of quantum mechanics, such as Niels Bohr, took this for granted (see \citealp{652}, for a critical discussion of Bohr's views on quantum mechanics, also in relation to modern takes on the problem of interpreting the formalism of this theory). In Bohr's words:


\begin{quote}
In [the context of quantum mechanics], even the old question of an ultimate determinacy of natural phenomena has lost its conceptional basis [...]\\
\citep[][p. 317]{432}
\end{quote}

 
On this basis, even modern metaphysicians may defend the claim that a scientifically informed doctrine of metaphysical realism has to let go of the requirement of definiteness (see e.g., \citealp{690} for a recent review of quantum indeterminacy). 

However, this plain reading of the quantum formalism raises some major conceptual questions. Most notably, one may ask why any quantum measurement always select just \emph{one} definite outcome out of the many possible outcomes of a coherent superposition. According to quantum mechanics, the answer is given by the collapse postulate: every time a system undergoes a measurement procedure, the state vector collapses into one of the eigenstates of the measured observable. However, this is not a satisfactory answer for the simple reason that the theory is silent about which sort of interactions must be described as ``measurements'' (and hence described by the collapse) and which as normal physical interactions between systems (described by Schr\"odinger's dynamics). In the absence of a precise criterion to discriminate between the two, the choice is usually left to practical considerations, which is sufficient for experimental purposes, not for philosophical ones. This is, in a nutshell, the famous \emph{measurement problem} of quantum mechanics. 

Clearly, for a scientific realist, quantum mechanics is not just a theory about measurements, it is a theory about the physical world. Hence, from a realist point of view, the measurement problem is an issue concerning the physical world itself: how can the classical features of the macroscopic world, such as definite objects having definite locations, be accounted for by a theory that extends quantum superpositions also to the macroscopic level via the Schr\"odinger evolution? This is usually the point where a ``standard realist interpretation'' of the quantum formalism is invoked.

To clarify what we mean by ``standard realist interpretation'', let us first of all consider the analysis of the measurement problem put forward in \cite{197}. Maudlin casts the measurement problem in terms of the conceptual tension among the following three statements: (a) quantum measurements always yield a unique definite result, (b) quantum states are informationally complete, and (c) the dynamics of the quantum states is encoded in the Schr\"odinger equation. Under the light of this analysis, the first step in providing  a standard realist interpretation of quantum mechanics is to give reasons to let go of one of the above statements. For example, Everettians do away with (a), Bohmians reject (b), while GRW theorists deny (c). 


The three mentioned interpretations are realist insofar as they maintain a \emph{sui generis} and robust notion of definiteness and objectivity. For example, the de Broglie-Bohm theory and GRW theory can be associated to a view in which the world is made up of an objective global arrangement of \emph{something} with a definite location evolving in time.\footnote{The way this may be implemented wildly varies in the literature. In the case of the de Broglie-Bohm theory, for example, one might have material particles in ordinary $3$-space plus quantum potential \citep{360}, material particles and a multi-field in $3$-space \citep{689}, material particles in $3$-space \emph{simpliciter} \citep{428}, or even a single world-particle and a local field in a $3N$-dimensional space \citep{376}. The important point is that each of these ontological pictures consists of entities with definite and objective features.} Instead, the Many Worlds interpretation is compatible with the view that there objectively is a wave function of the universe with a definite branching structure (cf. \citealp[][chapter 3]{668}). Hence, there is a clear way to preserve standard realism under the light of quantum physics: just acknowledge that the usual textbook reading of the quantum formalism---that is, assuming a unitary physical evolution for (closed) quantum systems, plus a wave function collapse upon measurement---is prone to the measurement problem, and subsequently adopt a standard realist interpretation that solves it.

Is this the only way to be a realist with respect to quantum mechanics? Many authors would answer this question in the negative. For these ``non-standard'' realists, the notion of standard realist interpretation delineated above perverts the true spirit of quantum mechanics in that it introduces an ``extra baggage'' of physical and/or philosophical assumptions. For example, according to Crull:

\begin{quote}
Historically, Bohmian mechanics and collapse interpretations were developed precisely in order to explain the appearance of classical phenomena (definite outcomes being among these). [...] [I]f one focuses on the specific problem of outcomes as the source of the anxiety, then the same is true of these interpretations as of many Everettian explanations: they supplement the standard formalism in order to get answers. In the case of Everett, this supplement is often philosophical but sometimes carries inescapable physical ramifications; in the case of Bohm and collapse interpretations, one must certainly admit additional physics alongside one's philosophy, e.g., the quantum potential or a non-unitary collapse mechanism. But if important theoretical work can be done without going these extra steps in any given direction, don't we stand to gain from the sustained generality of such an approach?\\
\citep[][pp. 1029-1030]{416}
\end{quote}

However, the question now arises as to how to consider quantum mechanics as a guide to shape metaphysical realism, while at the same time rejecting the standard realist interpretations (i.e., de Broglie-Bohm, GRW, and Many Worlds). A possible solution may be to sidestep the need for a standard realist interpretation by claiming that the measurement problem is not that big of a problem, after all. Indeed, for Crull, thinking that the measurement problem really is a problem betrays ``[...] a deep discomfort with accepting that quantum mechanics reveals a truly indeterministic world---a world that does not contain a causal (or any other) story about the `choice' of one approximately non superposed state over other approximately non-superposed yet equiprobable states. If physics has taught us that the world is indeterminate, then an answer to the specific problem of outcomes might well lie outside the scope of what is accessible or demonstrable.'' (\emph{ibid.,} page 1027). In the same vein, other authors like \v{C}aslav Brukner, go as far as claiming that ``the solution [to the measurement problem] lies in understanding that `facts' can only exist relative to the observer.'' \citep[][p. 95]{660}. This latter point is also hinted at in \cite{669}:

\begin{quote}
[T]he problem of the interpretation of quantum mechanics has not been fully disentangled. This unease, and the variety of interpretations of quantum mechanics that it has generated, are sometimes denoted the ``measurement problem.'' [...] [T]his unease may derive from [...] the concept of observer-independent state of a system, or, equivalently, the concept of observer-independent values of physical quantities.\\
(\emph{ibid.}, p. 1639)
\end{quote}

Clearly, questioning the requirements of definiteness and objectivity underlying standard realism is not enough to establish a different kind of metaphysical realism in the image of quantum mechanics. In order to do so, it is necessary to come up with a story about how the appearance of a classical world arises, which relies on the standard quantum formalism alone---i.e. without the additional physical and philosophical baggage that the standard realist interpretations carry. This is the point where decoherence appears in the debate (see, e.g., \citealp{632}, and references therein; \citealp[][section 1.2]{696}).


The goal of the present paper is to investigate the type of realism that best suits the framework of decoherence taken at face value without postulating a plurality of worlds, or additional hidden variables, or non-unitary dynamical mechanisms. We will call this attitude ``decoherence without interpretation'', to highlight the fact that it consists of a much more conceptually minimal realist take on the quantum formalism than the standard realist interpretations outlined above. We will firstly provide a very quick introduction to decoherence and its relation to the measurement problem (section \ref{sec:dec}). Sections \ref{sec:insig} and \ref{sec:cosm} will then focus on the relation between decoherence theory and metaphysical realism under the light of the minimal ontic commitments entailed by the ``decoherence without interpretation'' attitude.

\section{A short guide to decoherence}\label{sec:dec}

Decoherence is the theory of open quantum systems, i.e., systems interacting with each other and forming as a consequence new entangled states. Usually the decoherence mechanism involves two ``actors'', namely, a \emph{system} and an \emph{environment}. In general, the environment is simply defined as the collection of certain degrees of freedom that we decide to ignore in the description of the system but that still have some effect for the behavior of that system. Usual environments are, e.g., air molecules scattering off a table, or spinorial degrees of freedom relative to the electron position. Decoherence is, then, the physical effect one can measure on a subsystem of the total entangled state, namely, the loss of coherence between relative states of the initial superposition. With this respect, it is interesting to ask what is exactly achieved by decoherence and to what extent it explains (or helps explaining) the appearance of a determinate physical state at the macroscopic level.



\subsection{Density matrices}\label{sec:dens}

In this section, we introduce the principal mathematical tool used by decoherence, that is, the \emph{density matrix}. In decoherence theory, we are interested in the description of a subsystem of a larger entangled state (usually, a macroscopic object entangled with the external environment). However, in quantum mechanics, we cannot assign a state vector (or a wave function) to such subsystems. Nevertheless, they can be described
 by a specific type of density matrix, called \emph{reduced density matrix} (RDM). Understanding the physical meaning of the RDM is thus essential to understand the physical and philosophical implications of decoherence theory.


\subsubsection{Pure state density matrix}
Consider a quantum system represented by the pure state $\ket{\psi}$. The system density matrix 
$\hat{\rho}_{\psi} $ is, by definition, the projection operator into the state  $\ket{\psi}$:

\begin{equation}
\hat{\rho}_{\psi}= \ket{\psi}\bra{\psi}.
\end{equation}



Perhaps not surprisingly, the density matrix takes a particularly clear meaning when expressed in the matrix form. For simplicity, we consider as a pure state the superposition of two states : $\ket{\psi}=c_{\alpha}\ket{\alpha}+c_{\beta}\ket{\beta}$, where $ \ket{\alpha} and \ket{\beta} $ are eigenstates of a given observable $\hat{O}$, and $|\alpha|^2+|\beta|^2=1$. The matrix representation will be:\footnote{\label{diagz} The form of the density matrix \eqref{eq:pur} is basis-dependent and it is always possible to find a basis in which this matrix is diagonal. However, in most cases the bases that diagonalize a pure-state density matrix are not associated with physically meaningful observables.}

\begin{equation}\label{eq:pur}
\hat{\rho}_{\psi} =
\left( \begin{array}{ccc}
|c_{\alpha}|^2 & c_{\alpha}c_{\beta}^* \\
c_{\alpha}^*c_{\beta} & |c_{\beta}|^2  \\
\end{array} \right).
\end{equation}

The diagonal elements of the matrix represent the probabilities of finding the system in the states, respectively, $\ket{\alpha}$ or $\ket{\beta}$ if we perform a measurement of the observable $\hat{O}$.\footnote{Note that the probability distribution is always given by Born's rule.}
The off-diagonal elements of the matrix indicate, instead, quantum interference between the two definite states $\ket{\alpha}$ and $\ket{\beta}$.

The density matrix of a system is essentially a mathematical tool that describes all possible measurement outcomes and their probability distribution. This is also intuitively indicated by the \emph{trace rule}:\footnote{\rom{See e.g. \citet[][pp. 35-36]{431} for a derivation of this rule.} }



\begin{equation}\label{eq:trule}
Tr(\hat{\rho}\hat{O})=\braket{\hat{O}},
\end{equation}

which connects the mathematical procedure of the trace of the operators $\hat{\rho}\hat{O}$ with the physical concept of the mean value of the observable $\hat{O}$. 

\subsubsection{Reduced density matrix}\label{rux}
Finally, we consider the case in which the system is an entangled state of two subsystems: $\ket{S}$ and $\ket{E}$:




\begin{equation}\label{eq:SE}
\ket{\Psi}=a\ket{S_1}\ket{E_1}+b\ket{S_2}\ket{E_2}.
\end{equation}

This is the typical case in decoherence theory, where $\ket{S}$ is usually the macroscopic system under analysis and $\ket{E}$ the ``environment'', i.e. an external system interacting with $S$ through a suitable interaction Hamiltonian $H_{int}$, responsible for the entangled state between the two \rom{in the following way}:

\begin{subequations}\label{eq:int}
\begin{equation}
\ket{S_1}\ket{E}\stackrel{H_{int}}{\rightarrow}\ket{S_1}\ket{E_1},
\end{equation}
\begin{equation}
\ket{S_2}\ket{E}\stackrel{H_{int}}{\rightarrow}\ket{S_2}\ket{E_2}.
\end{equation}
\end{subequations}

In the entangled state \eqref{eq:SE}, it is not possible to assign an individual state to the subsystems $\ket{S}$ and $\ket{E}$. The only way to extract information about the possible measurement outcomes of one subsystem (for example, the subsystem $S$) is (i) to define a density matrix for this subsystem, and (ii) to compute the mean value of the observable via the trace rule.  The density matrix of the subsystem of a larger entangled state is the RDM, and is computed by tracing out the  degrees of freedom external to the considered subsystem---for example, by tracing out the ``environmental'' degrees of freedom of the subsystem $\ket{E}$ if we are interested in the RDM of the subsystem $\ket{S}$. In the matrix representation, it takes the following form:





\begin{equation}\label{eq:RDM}
\hat{\rho}_S^{red} =
\left( \begin{array}{ccc}
|a|^2 & ab^*\braket{E_2|E_1} \\
a^*b\braket{E_1|E_2} & |b|^2  \\
\end{array} \right).
\end{equation}

As in the case of the pure state density matrix, the diagonal elements represent the probability of finding the subsystem $S$ in the definite states $\ket{S_1}$ or $\ket{S_2}$ and the off-diagonal elements indicate quantum interference between those states. Using the RDM we can finally compute the statistical distribution of all possible outcomes of a measurement performed on the subsystem $S$:

\begin{equation}
Tr(\hat{\rho}_S^{red}\hat{O}_S)=\braket{\hat{O}_S}.
\end{equation}

We note that the RDM does not describe the physical state of the subsystem, since this information is encoded in the state vector of the system (however, this point may be controversial and it is not supported by density matrix realists, such as \citealp{691,692,693}). After the interaction, the two initial systems $S$ and $E$ do not possess anymore an individual state, but they are fused together in the new entangled state $\ket{\Psi}$.

\subsection{The decoherence process}\label{sec:proc}
From \eqref{eq:RDM}, we note that if the relative environmental states are orthogonal:
\begin{equation}\label{eq:robust}
\braket{E_i|E_j}=0,
\end{equation}
then the RDM of the subsystem becomes diagonal:

\begin{equation}\label{eq:RDM2}
\hat{\rho}_S^{red} =
\left( \begin{array}{ccc}
|a|^2 & 0\\
0 & |b|^2  \\
\end{array} \right).
\end{equation}
Eq. \eqref{eq:RDM2} indicates that if we perform a local measurement on the subsystem $S$, we do not detect quantum interference between the relative states $\ket{S_1}$ and $\ket{S_2}$. 
The loss of coherence between relative states of the subsystem due to a continuous interaction with the environment is the \emph{decoherence process}.\footnote{The diagonal form \eqref{eq:RDM2} achieved via environmental decoherence always selects physically relevant observables, contrary to the diagonalizations mentioned in footnote \ref{diagz}.}

A couple of remarks are in order at this point. First of all, the loss of coherence entailed by the decoherence process is only local: the phase relations between the relative states are still there, but ``hidden'' at the level of the entangled state (i.e., the phase relations are not destroyed but ``delocalized''). Secondly, even if decoherence transforms the initial coherent superposition into an incoherent (or improper) mixture of states, it remains rather silent about the physical state of the newly formed subsystem, as we cannot assign to it a state vector or a wave function (see \citealp{721}, for a detailed discussion of this point). What decoherence certainly does, however, is to change the empirical predictions of measurements performed on the subsystem of the entangled state. In fact, after the decoherence process, it is impossible to detect quantum interference between relative states of the substystem. 
And this is exactly what decoherence eventually  achieves: it makes subsystems \emph{look} classical if we measure certain observables or properties on the subsystem. Think for example of the double-slit experiment with an electron beam. If we perform the experiment in an isolated setting, an interference pattern will eventually be formed on the screen. However, if we let air molecules fill the space between slits and screen, the interference pattern will disappear and the final distribution of dots on the screen will be indistinguishable from the classical one because of the decoherence effect induced by the air molecules. However, this does not mean that the electron beam has become classical: in fact, we cannot assign definite trajectories to the electrons.

\subsection{The Schr\"odinger cat decohered}\label{sec:deco}
To make a concrete case, it may be useful to analyze the famous Schr\"odinger cat paradox from the point of view of decoherence. Suppose that the state of the radioactive atom \rev{in the box} is described by the superposition

\begin{equation}
\ket{A}=\frac{1}{\sqrt{2}}(\ket{A_D}+\ket{A_{ND}}),
\end{equation}
where the states $\ket{A_D}$ and $\ket{A_{ND}}$ represent, respectively, the ``decayed'' atom and ``not decayed'' atom. Following the causal chain described by Schr\"odinger (radioactive atom, Geiger counter, relais, hammer, poison, cat), the state of the atom gets eventually entangled  with the state of the cat in the box, leading to the famous coherent superposition of ``atom + cat'' states:

\begin{equation}\label{eq:caz}
\ket{A,C}=\frac{1}{\sqrt{2}}\ket{A_D}\ket{\ldots}\ket{C_D}+\frac{1}{\sqrt{2}}\ket{A_{ND}}\ket{\ldots}\ket{C_A},
\end{equation}
 where the states $\ket{C_D}$ and $\ket{C_A}$ represent, respectively, the states of the ``dead cat'' and ``alive cat''.

Similarly to the generic system discussed in the previous section, this is an entangled state of two subsystems, i.e., the subsystems ``atom'' and ``cat'' (the states correlated with them in between can be ignored). We are now \emph{free to consider} one subsystem as the system of interest and the other as the environment. Since we are interested in the behavior (and destiny) of the cat, we decide to select the ``atom'' as the environment and the ``cat'' as the system of interest. 

If the relative states of the atom are orthogonal, i.e., $\braket{A_D|A_{ND}}=0,$ the RDM of the subsystem ``cat'' is:
 
 \begin{equation}
 \hat{\rho}_C^{red}=\frac{1}{2}\ket{C_D}\bra{C_D}+\frac{1}{2}\ket{C_A}\bra{C_A},
 \end{equation}
 or, in the matrix form (with the two states ``dead cat'' and ``alive cat'' as basis states of the matrix):
 
\begin{equation}\label{eq:cat}
\hat{\rho}_S^{red} =
\left( \begin{array}{ccc}
\frac{1}{2} & 0\\
0 & \frac{1}{2}  \\
\end{array} \right).
\end{equation}

\eqref{eq:cat} tells us that, if we perform a measurement on the subsystem ``cat'', for example by opening the box and looking at the actual state of the animal, we have probability $0.5$ of seeing a definite ``dead cat'' state and probability $0.5$ of seeing a definite ``alive cat'' state. 
However, \emph{these probabilities are not epistemic}, since the two components of \eqref{eq:cat} refer to an improper (non-classical) mixture of states. That is, decoherence does not single out one of the components of the superposition. This means that the cat has not suddenly become a classical object: it is still a superposition of a definite ``alive cat'' and a definite ``dead cat'', even though we cannot detect interference between these two states. This is basically the reason why decoherence does not solve the measurement problem: it does not explain why we get to see one particular outcome and not the other \rev{(cf. \citealp{349}, and references therein)}.\footnote{In the literature, there is also a stronger objection to decoherence due to \cite{523}. In short, these authors claim that the decoherence program is circular because it assumes the collapse postulate and the Born rule to make physical sense of the diagonalization process \eqref{eq:RDM2}. An analysis of Okon and Sudarsky's argument falls outside the scope of the present paper. For the time being, we note that we disagree with their conclusion. As it is shown by the presentation above, all the tools of decoherence are consistently used, and a circularity would emerge only if one assumed that the RDM is able to provide one definite outcome---which is not how the RDM is standardly conceived of.} This is also the reason why it cannot explain (alone) the appearance of the classical world, for the solution of the measurement problem is a necessary condition to derive the appearance of macroscopic objects with well-defined positions and well-defined classical properties. This means that decoherence in and of itself cannot \emph{replace} the physical and/or philosophical extra baggage usually involved in reconciling quantum mechanics and scientific realism. Therefore, those scientific realists who are not sympathetic to the standard realist interpretations of quantum mechanics can \emph{at most} supply a set of assumptions that is different---more minimal, if you want---than those underlying these realist interpretations (more on this in sections \ref{sec:insig} and \ref{sec:cosm}), but they cannot just invoke decoherence disjointed from any interpretive move whatsoever (this point is stressed in \citealp{439}, and \citealp{671}).

\section{Metaphysical Perspectives on Decoherence}\label{sec:insig}
In section \ref{sec:deco}, we have argued that the ``decoherence without interpretation'' attitude still requires some interpretive move in order to be compatible with scientific realism. So, how exactly one should read off ontological morals from decoherence without resorting to any extra conceptual baggage required by the standard realist interpretations?

The first step in this sense might be to argue that the RDM is not just a computational tool, but should be granted ontological dignity. To this extent, one might point out that the very fact that the RDM gives us the right probabilistic distributions of measurement outcomes entails that it captures some underlying features of reality. Decoherence, then, would concern the dynamical behavior of such features. This is clearly an extra philosophical assumption put on top of the formalism, since the decoherence formalism alone does not treat the RDM on a par with a genuine quantum state. Therefore, it is clear that ``decoherence without interpretation'' still is some sort of non-standard, minimal realist interpretation of the quantum formalism. But what are the features of reality captured by the RDM, according to this minimal interpretation?

To answer this question, we should start by considering how the decoherence process impacts on the picture of reality of someone observing the decohered system. Let's consider two different observers in the cat's scenario described by the RDM given in \eqref{eq:cat}. Assume that one of the observers---let's call him $A$---opens the box with the cat while being isolated in a room, with the second observer---$B$---waiting outside. Now, for $A$ the cat will be, say, alive with certainty,\footnote{This may raise the legitimate suspicion that a collapse mechanism is at work in the background. However, supporters of decoherence without interpretation would probably reiterate that there is just no fact of the matter accounting for the appearance of a definite outcome upon mesurement.} while $B$ would conclude that now the system is in a superposition of cat alive with $A$ seeing it alive and cat dead with $A$ seeing it dead. So one may ask: is the cat alive (as seen by $A$) or neither definitely alive nor definitely dead (as described by $B$)? Worse still, if at this point $B$ enters the room, he has $0.5$ probability of ending up seeing $A$ observing a \emph{dead} cat. In short, it seems rather hard to come up with a coherent story that reconciles the ``reality'' experienced by $A$ with that experienced by $B$.


This issue is very-well known in the literature since the inception of quantum mechanics (the ``Wigner's friend'' thought experiment is probably the most mentioned source in this sense, see \citealp{662}). \cite{661}, for example, note that:\footnote{We have slightly modified the wording of the passage (not its meaning) to fit our example.}
\begin{quote}
``Objectively---that is, [\emph{for $B$}], who consider[s] as ``object'' the combined system [(box, $A$)]---the situation seems little changed compared to what [he] just met when [he was] only considering [the box]. The observer [$A$] has a completely different impression: for him it is only the [box] which belong[s] to the external world, to what he calls ``objective''. He attributes to himself the right to create his own objectivity [...] by declaring [...] ``I see [a live cat]''.\\
(\emph{ibid.}, pp. 251-252, emphasis in the original)
\end{quote}

This leads them to conclude that:
\begin{quote}
In present physics the concept of ``objectivity'' is a little more abstract than the classical idea of a material object. Is it not a guarantee of ``the objectivity'' of an object that one can at least formally attribute measurable properties to it in a continuous manner even at times when it is not under observation? The answer is No, as this new theory shows by its internal consistency and by its impressive applications.\\
(\emph{ibid.}, p. 259)
\end{quote}

More recently, \cite{658} and \cite{660} have devised variations of the ``Wigner's friend'' experiment in which a group of observers measuring the very same quantum system gets mutually inconsistent assignments of values to the measurement outcomes (see \citealp{659} for a nice exposition and a critical discussion of these arguments).

What kind of metaphysical realism can possibly go with such a highly non-objective characterization of reality? The answer, according to authors like \cite{653} is pretty straightforward:\footnote{Again, we have slightly changed the passage to fit our cat example.}

\begin{quote}
Our proposed perspectival way out of this dilemma is to ascribe more than one state to the same physical system. In the case under discussion, with respect to [$B$], representing the outside point of view, the contents of the laboratory room are correctly described by an entangled pure state so that we should ascribe improper mixtures (obtained by ``partial tracing'') to the inside observer [and the box]. But with respect to the inside observer (or with respect to the [box] in the room) the [cat is definitely alive (or dead)]. So the inside observer assigns a state to his environment that appropriately reflects this definiteness.\\
(\emph{ibid.}, p. 56)
\end{quote}


One might be tempted to identify Dieks' notion of ``perspective'' with that of ``Everettian robust branch''. This would be misleading because the latter notion refers to the different decoherent components of the RDM: In London and Bauer's experimental setup, this would translate into the fact that the inside observer $A$ splits into two copies, one observing the alive cat, and the other observing the dead cat. For Dieks, instead, the two perspectives are (i) that of  the inside observer $A$, and (ii) that of the external observer $B$: Both (i) and (ii) obtain in the same world, i.e., the observers do not split into different copies of themselves. 





Dieks' account of London and Bauer's thought experiment seems to point in the direction of some sort of \rev{strong variant} of perspectival realism. In a nutshell, this \rev{strong position would accept} that there is a mind-independent reality while denying that such a reality is an all-encompassing state of affairs which obtains independently of how it is observed. \rev{In the words of \cite{654}:}

\begin{quote}
\rev{My interest [...] is to explore the possibility of the potentially radical suggestion that perspectivalism can be extended to account for a type of objectivity, a type of observer-independence, in science, but where facts about scientifically modelled objects are nonetheless indexed to an observer perspective. My motivation is a recent set of claims from quantum foundations that quantum mechanics rules out the possibility of ``observer-independent facts''.\\
(\emph{ibid.}, pp. 2-3)}
\end{quote}

Here we are using the ``strong'' designation to distinguish this kind of perspectivism from the milder one advocated, e.g., in \cite{663,666}. This milder brand of perspectival realism, in fact, represents a weakening of the epistemic and semantic theses underlying scientific realism, but it fully preserves the metaphysical thesis of standard realism as we have set out in section \ref{sec:intro}. In other words, according to mild perspectival realism, even though \emph{there is} a mind-independent objective reality, there isn't a unique and objectively true scientific framework that grants full epistemic access to it (see \citealp[][section 1]{630}, for a concise statement along these lines). 

The strong variant of perspectivism suggested by the above quotations is instead a metaphysical thesis that radically departs from standard realism: even though there is a mind-independent reality, such a reality is not a coherent union of objective, that is, perspective-independent facts. This is because a fact is ``objective'' only with respect to a given perspective, which in turn implies that there is no all-encompassing picture of reality (a ``God's eye'' view of the world, so to speak).


Indeed, many other voices in the quantum foundations literature tend to sweep realism towards this strong perspectival direction.\footnote{Among them, it is worth mentioning \cite{673}, who however adopt a modal interpretation of quantum mechanics.} For example, \cite{656} writes:

\begin{quote}
Since the advent of quantum theory [...] there has always been a nagging pressure to insert a first-person perspective into the heart of physics. [The views acceding to this pressure] have lately been termed ``participatory realism'' to emphasize that rather than relinquishing the idea of reality (as they are often accused of), they are saying that reality is more than any third-person perspective can capture.\\
(\emph{ibid.}, p. 113)
\end{quote}

The umbrella term ``participatory realism'' collects all the realist and single-world interpretational frameworks that rely on the assumption that, in any given situation, there is no unique quantum state associated to a system, but an observer-dependent plurality of them. Such a characterization makes it crystal clear that participatory realism bears an essential perspectival component with it. Three important examples: \cite{695} has recently argued that Qbism can be considered a perspectival normative realist stance; \cite{696} defend the view that quantum mechanics is fundamentally about relative facts, some of which become stable in light of decoherence; \cite{697} frame quantum measurements as the realization of a given definite value of the observable depending on the measurement context.

We have finally got a clear metaphysical sense of the direction in which philosophers who support decoherence without interpretation are heading when they claim that we should conceive of the quantum reality as non-classical. Simply speaking, their attitude nicely fits a conceptual framework in which the picture of reality entailed by decoherence trades the usual requirement of (global) objectivity for a perspective-indexed (or context-indexed) type of objectivity. 

It might be claimed that global objectivity is not really lost on this view, but has just to be pragmatically interpreted as ``intersubjective agreement'' (see, e.g., \citealp{700}, for a discussion of this type of objectivity). That is, when a sufficient number of observers agree on a given fact (e.g., regarding a measurement outcome), then such a fact is objective for all practical purposes. While such a move may be helpful in salvaging some conspicuous aspects of epistemic and semantic objectivity, it cannot make up for a notion of \emph{metaphysical} objectivity because it does not concern the way things are---it is in fact a \emph{pragmatic} notion. Hence, it does not seem that invoking intersubjective agreement can make the rejection of global objectivity a less radical move.

Historically, the first attempt to define  objectivity as a kind of intersubjective agreement about the state of a system in the context of decoherence is due to \cite{701} through the approach called \emph{quantum Darwinism}. This approach starts from the notion of \emph{einselection}, which amounts to the selection of pointer states by the environment. In a nutshell, einselection describes which particular states of a system will survive the decoherence process---the ``survival of the fittest''. 
Einselection, however, does not select a \emph{unique} definite state (i.e., it does not get rid of the quantum superposition), so it provides no explanation as to why the system can yield a definite measurement outcome recorded by many independent observers. According to quantum Darwinism, this illusion of objectivity is due to the redundancy of copies of certain relative states of the system in the environment. These redundant---or, \emph{robust}---states are those that do not change under the action of the environment. As a consequence, they get entangled with many relative states of the environment, thus forming a multitude of copies that spread out in the surrounding environment. In other words, when a certain state is robust, large ``fragments'' of the environment will spread the information about this state: the observers will just intercept these fragments and acquire information about the system. The more the information about the system state via fragments is redundant in the surrounding environment, the more the different observers will acquire information about the state without destroying it (as the information is brought about by independent copies of the environment). As the very same information about the decohered state is transferred to the majority of the observers, that state---according to Zurek---will become objective, in the sense that all observers will have a consensus on the state of the system, despite the underlying superposition. In Zurek's words:

\begin{quote}
The proliferation of records allows information about [the system] $\mathcal S$ to be extracted from many fragments of [the environment] $\mathcal E$ [...] Thus, $\mathcal E$ acquires redundant records of $\mathcal S$. Now, many observers can find out the state of $\mathcal S$ independently, and without perturbing it. This is how preferred states of $\mathcal S$ become objective. Objective existence---hallmark of classicality---emerges from the quantum substrate as a consequence of redundancy.\\
(\emph{ibid.}, p. 182)
\end{quote}


In this way, quantum Darwinism accounts for objectivity as the intersubjective agreement about the system's state. This is probably the first explicit attempt to define a kind of participatory realism in the context of standard decoherence to explain the (appearance of) stability, uniqueness and classicality of the pointer states selected by the environment.\footnote{Quantum Darwinism is conceptually close to the strategy recently developed in \cite{696}, according to which decoherence tends to promote a restricted collection of relative states to robust states, i.e. states that are equivalent for most observers. With this respect, Di Biagio and Rovelli's strategy can be thought of as a refinement of Zurek's quantum Darwinism.}


To sum up, taking decoherence at face value without resorting to a standard realist interpretation leads to a rejection of standard objectivity, which has to be put on top of the withdrawal of ontic definiteness mentioned in section \ref{sec:intro}. This obviously represents an extreme reshaping of the metaphysical component of scientific realism. However, some may object that this ``decoherence-induced'' perspectival drift of scientific realism can hardly be called ``scientific realism'' anymore given that it either gives up or radically revises (e.g. by making them perspective-dependent) the three theses on which scientific realism is built (see, e.g., \citealp{665}, for an attack on perspectivism broadly construed). Here we are not interested in pursuing this line of argument. Instead, we want to explore some conceptual consequences of accepting this strong perspectival realism. This will be done in the next section.

\section{A further weakening of reality: decoherence in cosmology}\label{sec:cosm}
\rev{In section \ref{sec:deco}, after presenting the (``atom'',``cat'') entangled state \eqref{eq:caz}, we have mentioned the fact that we were quite free to choose which degrees of freedom to trace out in order to construct a RDM for the subsystem of interest. This is a simple example that vividly shows how the distinction between system and environment is not strictly speaking built in the decoherence process.  In other words, in decoherence theory the division between certain degrees of freedom that collectively we call the ``system'' and other degrees of freedom that we call ``the environment'' has always a certain degree of arbitrariness.\footnote{\rev{The problem of individuating quantum subsystems was well-known even before the decoherence formalism made its inception. For example, \citet[][chapter VI]{200}, was one of the first to attempt at providing a characterization of the division between measured system and measuring apparatus.}} In standard cases, however, this choice is in some sense a ``given'' to physicists for the simple reason that the decoherence process starts with an interaction between two \emph{distinct} systems initially unentangled (see \eqref{eq:int}).}

However, we want to argue that there is a point after which \rev{drawing} the system/environment distinction becomes so arbitrary to impact on the (metaphysical) perspectives generated by the decoherence process. This point of no return is marked by quantum cosmology.


An important caveat is in place before discussing the matter further: simply speaking, there is no well-established theory of quantum cosmology as yet. What we have instead is a (fractured) collection of physical models, usually involving the framework of quantum field theory over a curved spatiotemporal background, which aim to describe a number of aspects of the origin and evolution of the universe that cannot be accounted for by classical cosmology (which in turn is based on general relativity). In this sense, we are not claiming that the considerations we made earlier about decoherence smoothly carry over from standard, non-relativistic, quantum mechanics to quantum cosmology: probably they will once a complete theory of quantum cosmology will be established, but we are not there yet. That being said, we anyway believe that (i) the results achieved so far in the context of quantum cosmology, however partial they might be, are still physically relevant enough to be investigated from a metaphysical perspective and (ii) these results rely on a core conceptual framework which is shared with non-relativistic quantum mechanics. For these reasons, we think that, even at this stage of theoretical development, it makes a lot of sense to discuss how a conceptual analysis of decoherence translates to the cosmological context.

Let us now focus on a concrete physical issue investigated in quantum cosmology, which has been brought up several times in the philosophical debate about decoherence mentioned earlier. This issue concerns the formation and subsequent evolution of the material structures of the universe from an early inflationary epoch, which left the universe in a homogeneous and isotropic state. Roughly speaking, the question is: how did the ``messy'' universe we observe at all scales pop out of the perfectly ordered one that was left after the (hypothesized) sudden expanding phase physicists refer to as ``inflation''? This question is worth being asked because the inflationary model of the universe is very valuable in that it is able to provide convincing solutions to many problems that arise in classical cosmology. For example, the highly ordered state of the universe after inflation explains the observed homogeneity of the cosmic microwave background. Moreover, the quantum fluctuations of the inhomogeneous terms of the inflaton field show a striking similarity with the power spectrum of the observed density fluctuations that can be traced back to the early stages of the universe's expansion. Hence, it makes sense to see what kind of story, if any, the inflationary model provides for the origin and development of these so-called ``seeds'' of cosmic structure. The following sketch is based on the presentation given in \citet[][section 2]{629} and \citet{628}, and it has no pretense of formal rigor beyond that required to drive home our philosophical point.

The simplest model\footnote{In the following, we set $c=G=\hbar=1$.} starts from a (classical) scalar field $\phi$ coupled to the metrical field $\mathbf{g}$ of general relativity.  $\phi$ is called the \emph{inflaton} field because it obeys some appropriate dynamic constraints, which assure that the universe undergoes a sudden accelerated expansion in its early stages. The model focuses on small perturbations $\delta\phi$ of the inflaton field while holding fixed the spacetime geometry $\mathbf{g}$, which hence acts as a background against which these perturbations propagate. Usually, these perturbations are treated as a \emph{sui generis} scalar field $y(\mathbf{x},t)=a(t)\delta\phi(\mathbf{x})$, $a(t)$ being the scale-factor of the universe, which changes in cosmic time $t$. In the Big Bang scenario, $a$ is a monotonically increasing function such that a(0)=0. This new field is then decomposed in its Fourier modes:

\begin{equation}\label{eq:flu}
y(\mathbf{x},t)=\int\frac{d^3 k}{(2\pi)^{\frac{3}{2}}}y_{\mathbf{k}}(t)e^{i\mathbf{k}\cdot\mathbf{x}}.
\end{equation}

We can then exploit the resources of analytic mechanics to calculate the momenta conjugate to these modes (we suppress the time dependence for notational simplicity):

\begin{equation}\label{eq:pcl}
p_{\mathbf{k}}=y'_{\mathbf{k}}-\frac{a'}{a}y_{\mathbf{k}},
\end{equation}
where a prime symbol indicates a derivative with respect to cosmic time. 

The appropriate general relativistic action for the system under consideration yields the equations of motion for the Fourier modes:

\begin{equation}\label{eq:mod}
y''_{\mathbf{k}}+\left (k^2 -\frac{a''}{a}\right )y_{\mathbf{k}}=0,
\end{equation}
with $k=|\mathbf{k}|$.

When this model is canonically quantized, both $y_{\mathbf{k}}$ and $p_{\mathbf{k}}$ are promoted to operators, and from them, the annihilation and creation operators ($\hat{\mathfrak a}_{\mathbf{k}}$ and $\hat{\mathfrak a}^{\dagger}_{\mathbf{k}}$, respectively) can be defined in the usual way.

The initial vacuum state $\Psi_0(t_0)$, i.e., the one such that $\hat{\mathfrak a}_{\mathbf{k}}\Psi_0(t_0)=0$ for all $\mathbf{k}$, corresponds to a quasi-Minkowski vacuum and, hence, is spatially homogeneous and isotropic. It can be shown that $\Psi_0(t_0)$ is the minimum uncertainty wave packet, that is, a Gaussian state.

The quantum dynamical evolution of the model is such that, at the onset of inflation, the wavelengths of the fluctuations are much smaller than the curvature scale (i.e., $k^2\gg\frac{a''}{a}$). Soon, however, some wavelengths grow much larger than the curvature scales and the corresponding modes become ``squeezed''. Roughly speaking, squeezing happens when pairs of particles with opposite momenta are created. A squeezed mode corresponds to a Wigner quasiprobability distribution over phase space which is basically $|\Psi_0|^2\delta \left(p_{\mathbf{k}}-p_{cl}(y_\mathbf{k})\right)$, where $p_{cl}$ is the classical momentum conjugate to $y_\mathbf{k}$ (see equation \eqref{eq:pcl}). This means that the eigenvalues of $y_\mathbf{k}$ corresponding to squeezed modes follow Born's rule as usual, but to each of them is associated a definite value of the conjugate variable $p_{\mathbf{k}}=p_{cl}(y_\mathbf{k})$. This is taken as a justification of the fact that the modes with wavelength such that $k^2\ll\frac{a''}{a}$ can be considered equivalent to a statistical ensemble of classical fields. Once this step is achieved, it is just a matter of mathematics to show that the expectation values for these squeezed modes are equivalent to the averages of classical random variables, from which it is possible to establish a link between the ``primordial'' quantum fluctuations and the classically behaving inhomogeneities observed in the cosmic microwave background (which are the ``seeds'' of cosmic structure mentioned above).

Now it seems that we have been able to kill two birds with one stone, namely, to provide a quantum-to-classical transition mechanism \emph{and} to explain how a structured universe originated from the quantum fluctuations of an early highly symmetric stage. However, it is too soon to cry victory. In fact, the above treatment has focused on a small part of the whole quantum state, i.e., squeezed modes, leaving conveniently aside all the other modes. It is exactly at this point that the decoherence mechanism is invoked. Simply speaking, non-squeezed modes are regarded as negligible, and treated as (part of) the ``environment'' to be traced away from the RDM that accounts for the squeezed modes. But this begs the question as to what the ``source'' of entanglement between different modes of the same quantum state would be. The usual answer to this question is that a full quantum theory of the inflaton field will be non-linear, which will lead to the entanglement between modes corresponding to different wavelengths at different curvature scales. Another possible answer would be to claim that such a full quantum theory will feature other fields which might provide the environmental degrees of freedom needed for the decoherence mechanism to work. While both of these proposals are a viable solution to make decoherence work in this context as a computational tool, their implications from a scientific realist perspective are equally challenging, as we are going to argue below.

To see this, we have to go back to the issue of the distinction between system and environment. As we have said, in ordinary contexts the decoherence mechanism naturally induces a robust distinction between system and environment. For example, in the Wigner's friend scenario discussed in section \ref{sec:insig}, observers $A$ and $B$ certainly agree that the main system under investigation is a cat (although they disagree on the details of its physical description). In the cosmological case, instead, the situation is much more subtle. Indeed, in such a case, the system is the entire universe, which means that the kind of evolution \eqref{eq:int} is out of question because we do not have two initially distinct pure states getting entangled as a consequence of a physical interaction: there is (in a tenseless sense)  just a single quantum state of the universe.\footnote{An interesting analogy is drawn by \cite{686} between the case of quantum cosmology and the case of an atomic nucleus in nuclear physics. In both cases, a total symmetric/homogeneous state can be (effectively) decomposed via decoherence into a superposition of asymmetric/inhomogeneous states. In the example of the nucleus we have different definite orientations in space, while the total state of the nucleus is perfectly symmetric. The key insight that enables the analogy is that, in both cases, we deal with closed systems, and hence the environment is composed of internal degrees of freedom. This means that the effects of decoherence are seen from ``the inside'' of the system. We thank an anonymous referee for drawing our attention on this analogy.} \cite{655} nicely summarize this issue as follows:

\begin{quote}
If someone hands you two qubits $A$ and $B$, there is a well-understood procedure for constructing the quantum description of the composite system constructed from the two of them. If the individual Hilbert spaces are $\mathcal{H}_A  \simeq \mathbb{C}^2$ and $\mathcal{H}_B \simeq \mathbb{C}^2$, the composite Hilbert space is given by the tensor product, $\mathcal{H}\simeq \mathcal{H}_A \otimes  \mathcal{H}_B\simeq \mathbb{C}^4$, where $\simeq$ represents isomorphism. The total Hamiltonian is the sum of the two self-Hamiltonians, $\hat{H}_A$ and $\hat{H}_B$, acting on $\mathcal{H}_A$ and $\mathcal{H}_B$, respectively, plus an appropriate interaction term, $\hat{H}_{int}$, coupling the two factors. What about the other way around? If someone hands you a four-dimensional Hilbert space and a Hamiltonian, is there a procedure by which we can factorize the system into the tensor product of two qubits? In general there will be an infinite number of possible factorizations [...] Is there some notion of the ``right'' factorization for a given physical situation?
In almost all applications, these questions are begged rather than addressed.\\
(\emph{ibid.}, p. 3)
\end{quote}

The standard way out of this issue, as we have seen in the case of the seeds of cosmic structure, is to choose the decomposition of the ``universe'' system in relevant and irrelevant degrees of freedom depending on computational convenience, and show how the former decohere against the background of the latter. But this immediately prompts a key question: What are the metaphysical morals to be drawn from this move? Recalling what has been said in the previous section, one may want to argue that, behind the freedom to draw a line between system and environment in a cosmological context lies a further perspectival layer of reality. 

In the Wigner's friend case, in fact, the perspectival move was grounded in the claim that there is no objective fact of the matter about the cat's state, so each observer had its own perspective on that. By way of analogy, one may argue that the fundamental system/environment arbitrariness at a cosmic level implies that there is no fact of the matter whether there is a \emph{unique} division between system and environment and, hence, to each division corresponds a different perspective. Bringing back the Wigner's friend case, now observers $A$ and $B$ may also disagree on what the main system is even if they are interested in investigating the same physical situation (each of them can ``carve'' the universe in a different way depending on their computational needs). A moment of reflection, however, shows that this second perspectival layer is quite puzzling if analyzed through the lenses of standard realism. 


In fact, let's assume that it is possible to apply a \rev{strong} perspectival realist reading to cosmological decoherence, and let's go back to the inflationary picture sketched above. We have said that decoherence explains how the inhomogeneous seeds of cosmic structure popped out of an homogeneous quantum state. Hence we can claim that, from our perspective, the inhomogeneous structure of the universe is real. However, also from our perspective, we see that the cosmic background radiation is overwhelmingly homogeneous and isotropic, and we explain it by invoking the symmetry of the universal quantum state. Here there is a clear sense in which we speak about the ``same perspective'', namely, the fact that both observations are accounted for by the same observers using the \emph{same cosmological model}. However now we are in a strange situation: from our \emph{very same perspective}, it is true that the universal quantum state is \emph{both} symmetric \emph{and} inhomogeneous. Basically we let it undergo either the Schr\"odinger evolution or the non-unitary subsystem dynamics encoded in the decoherence process depending on what feature of \emph{our very same perspective} we want to account for. 

One may reply to this remark by pointing out that the total state was symmetric \emph{before} undergoing decoherence, but this is in stark tension with the fact that the universe does not interact with anything outside itself: the only thing that selects the inhomogeneous parts of the state at a later time is \emph{us}, and only for computational purposes. Recall from section \ref{sec:insig} that a perspective \emph{\`a la} Dieks is associated to a certain quantum state, and that a subsystem of a larger entangled state (formally represented by a RDM) cannot be assigned a physical state (end of section \ref{rux}). Hence, there is no question whether we can consistently assign different perspectives in the cosmological model under consideration.

In the end, it seems that the perspectival story applied to the system/environment distinction becomes extremely radical (at least, from the standard realist standpoint). Indeed, insisting on this story would mean claiming that mutually inconsistent facts obtain from the same perspective. Note how the strength of the above argument does not depend on the detailed form of the quantum cosmological state or on the particular coupling mechanisms underlying the decoherence process, but relies on a basic set of features of the model, i.e., closedness and global symmetry. From this point of view, the situation arises whenever we consider decoherence happening ``inside'' a closed system, that is, for situations in which the environment is composed of internal degrees of freedom of the total system.

However, those who defend a literal metaphysical reading of decoherence may retort by pointing out that there is no fact of the matter whether there is a unique division between system and environment at the cosmological level just because, in fact, \emph{there is no such thing as a system/environment divide simpliciter}. In this way, they would end up with an anti-realist attitude towards the mereological structure of the world,\footnote{See \cite{672}, for a presentation and defense of mereological anti-realism. Note that this position should not be confused with mereological nihilsm. This latter view maintains that there are only mereological simples, but it does not deny that the world possesses at least a weak mereological structure given by the whole/simples relation.} which is summarized in the following passage:

\begin{quote}
\rev{Ubiquitous and un controllable entanglement begets decoherence which in turn creates further entanglement: the system and environment become a new inseparable composite with a new environment to which it then becomes entangled, forming a new inseparable composite interacting with a still different environment, ad infinitum. The result is an upward cascade of entanglement relations largely hidden from experience by decoherence. In such a world there can be no material parts with true ontic grit. There can be no in-principle physical joints for carving nature, because any such joint is a mere specter whose existence and form depend entirely upon the specifics of environmental interactions.\footnote{This point is already present in \citet[][section 3, p. 881]{708}: ``Decoherence also ought to cure us of the notion that the world is truly divisible [...] One can slice the material world in myriad ways and get away with it because those slices have
no ontological grit [...]''.}\\
(\citealp{667}, p. 11)}
\end{quote}

Although this mereological anti-realist turn does not necessarily conflict with the scientific practice, it renders the whole metaphysical stance associated to decoherence even more radical. In fact, it demotes boxes, cats, observers, and so on to ``mere specters''. As a consequence, the mereological anti-realist has to do much more conceptual work than the standard realist in supplying a convincing story about why (i) it is physically meaningful to perform experiments on subsystems of the universe and (ii) we can rely on the conclusions drawn from these experiments in order to get a picture of the world.

To conclude, scientific realism has reached a crossroads. On the one hand, we can assume that the standard formalism of quantum mechanics---including decoherence theory---without any standard realist interpretation provides a complete description of reality, and accept the cost of a strong form of perspectival realism, possibly combined with mereological anti-realism. On the other hand, we can maintain that quantum mechanics and decoherence need a standard realist interpretation to provide a coherent story about the physical world and hence accept the extra baggage of parallel worlds, hidden variables, and objective collapses. We leave the final choice to the reader, being content to note that in the metaphysics of quantum mechanics there is definitely no free lunch.

\pdfbookmark[1]{Acknowledgements}{acknowledgements}
\begin{center}
\textbf{Acknowledgements}
\end{center}

We are grateful to Vera Matarese and two anonymous referees for their helpful comments. One of us (AV) acknowledges financial support from the Polish National Science Centre, grant nr. 2019/33/B/HS1/01772.

\pdfbookmark[1]{References}{references}
\bibliography{biblio.bib}
\end{document}